\documentclass[prl,twocolumn,showpacs,preprintnumbers,amsmath,amssymb]{revtex4}

\usepackage{graphicx}
\usepackage{dcolumn}
\usepackage{bm}

\begin{document}


\title{Effects of orientation and alignment in high-harmonic generation and above threshold ionization}

\author{C. B. Madsen}
\author{A. S. Mouritzen}%
\author{T. K. Kjeldsen}
\author{L. B. Madsen}\affiliation{%
Lundbeck Foundation Theoretical Center for Quantum System Research, Department of Physics and Astronomy, University of Aarhus, 8000 Aarhus C, Denmark.
}%

\date{\today}

\begin{abstract}
When molecules interact with intense light sources of femtosecond or
shorter duration the rotational 
degrees of freedom are frozen during the response to the
strong nonperturbative interaction. We show how the frozen degrees
of freedom affect the measurable signals in high-harmonic
generation and above threshold ionization.  High-harmonic
generation exhibits optical coherence in the signal from different
orientations of the molecule.  For ionization, the contributions from
different orientations are added incoherently. The effects are  demonstrated for realistic alignment and orientation schemes.
\end{abstract}





\maketitle

Currently, intense few-cycle pulses of femtosecond duration are
produced in several laser laboratories world-wide. It is of
fundamental interest to investigate how such pulses interact with
quantum systems. Gas phase atoms and molecules are ideal test
systems for these studies. As this research field is maturing,
interesting applications are evolving: tomographic
reconstruction~\cite{itatani04}, laser-induced electron
diffraction~\cite{Niikura02} and molecular
clocks~\cite{Niikura03,alnaser:183202,alnaser:163002,S.Baker04212006}
being examples under current study.

Molecules are particularly well-suited for such studies since they
inherently carry the time scales that match those of the new laser
sources: nuclei move on the femtosecond timescale, electrons on
the attosecond timescale. Hence, the new ultrashort sources are
perfect for looking inside molecules and for gaining insight into
nuclear and electronic dynamics. Conversely, if the dynamics of the
system under study is well-understood, information about the
characteristics of the few-cycle pulse can be obtained. For
instance, in atoms, where an accurate description of the electrons
can be obtained, the carrier-envelope phase difference may be
extracted in this way~\cite{paulus:253004,Haworth06,martiny:093001}.
For molecules, the extra nuclear degrees of freedom may be used to
control the process of interest. As an example, the ability to
orient molecules~\cite{stapelfeldt03} with respect to an external
axis may be used to enhance the high-harmonic
yield~\cite{PhysRevA.65.053805,itatani:123902,Kanai05,Lein05}.
Solving the dynamics of the molecule in the strong field
is, however, much more difficult than the atom due to the extra degrees of
freedom. In fact, not even the simplest case of H$_2^+$ interacting
with strong IR fields has been solved in full 6-dimensional time
dependent calculations. Fortunately, the interesting prospect
of ever shorter pulse durations introduces a simplification in the
description: the timescales of rotation and vibration are often much
longer than the actual applied pulses themselves, and therefore some
of these degrees of freedom may be treated as frozen during the
interaction with the field. In the present work, we describe how
frozen degrees of freedom affect the outcome of an experiment in a
non-trivial manner. We show that the influence of the dynamics and
the formulation of the theory of measurement depend very much on the
process considered: we obtain completely different behavior for
above threshold ionization (ATI) and high-harmonic generation (HHG)
with respect to the coherence in the signal from rotational degrees
of freedom. The discussion is exemplified using realistic alignment and orientation 
schemes.

We consider HHG and ATI in molecules interacting with an ultrashort
strong laser pulse. In these experiments, there are many molecules
in the laser focus, but the phase space density is low and we can
use the single-particle density operator $\hat{\rho}(t_0)$
for  calculations. At time $t = t_0$ prior to any probe or pump
pulse, the molecule is in a time-independent thermal state at
temperature $T$. By definition, $\hat{\rho}(t_0) =
\exp{(-\hat{H}/k_B T)}/Z$, with partition function $Z =
\textnormal{Tr}[\exp(-\hat{H}/ k_B T)]$, $\hat{H}$ the field-free
molecular Hamiltonian and $k_B$ Boltzmann's constant.
The initial state is resolved on energy eigenstates $|{\bm{\alpha}}\rangle$ with energy
$E_{\bm{\alpha}}$. We concentrate on diatomics
where, prior to the applied pulses, only the electric and
vibrational ground states are populated. Consequently, the energy eigenstate
is characterized by  the angular momentum quantum number, $J$, and its
projection on a space fixed axis, $M$, i.e., ${\bm{\alpha}}=(J,M)$. The discussion is
straightforwardly generalized to more complicated cases and the
conclusions remain unaffected. There is no decay on the timescales
we are considering so propagation is described by a unitary operator
$\hat{U}(t)$: $\hat{U}(t)|\bm{\alpha}\rangle =
|\Psi_{\bm{\alpha}}(t)\rangle$; $\hat{\rho}(t) = \hat{U}(t)
\hat{\rho}(t_0) \hat{U}^\dagger(t) = \sum_{\bm{\alpha}}
P_{\bm{\alpha}} |\Psi_{\bm{\alpha}}(t)\rangle
\langle\Psi_{\bm{\alpha}}(t)|$, with the Boltzmann weight
$P_{\bm{\alpha}} = \exp{(-E_{\bm{\alpha}}/k_{B}T)}/Z$. The evolution
due to $\hat{U}(t)$ can contain both alignment pulses and a
subsequent ultrashort probe pulse producing ATI and HHG.

We separate out the relatively slow rotational movement of the nuclei 
to obtain $\Psi_{JM}(\bm{r}_e,R,\Omega,t) \approx
\psi(\bm{r}_e,R,t;\Omega)\, \phi_{JM}(\Omega,t)$, with $R$  the
internuclear distance and $\Omega=(\theta,\phi)$ the spherical
polar solid angle composed of the usual polar and azimuthal angles.
The variables that enter only parametrically in the wave function are put after 
the semicolon. If we consider the response to a femtosecond probe pulse, the
rotational degrees of freedom can be considered frozen during the
probe pulse centered at $t_p$, i.e., the full wave function is approximated by
\begin{eqnarray}
\label{wavefunction} \Psi_{JM}(\bm{r}_e,R,\Omega,t) \approx
\psi(\bm{r}_e,R,t;\Omega,t_p)\, \phi_{JM}(\Omega,t_p).
\end{eqnarray}
 In the case of a prealigning or orienting pump pulse between $t_0$ and $t_p$,  $\phi_{JM} (\Omega, t_p)$ is the rotational wave packet evolving from $Y_{JM}(\Omega)$ at time $t_0$. If no pump pulse is used then $\phi_{JM} (\Omega,  t_p) =
 Y_{JM}(\Omega)$.

Treating HHG first, the complex amplitude for the emission of
harmonics polarized along the unit vector $\bm{e}$ with frequency
$\omega$, is obtained from the Fourier transform of the dipole
acceleration
\begin{equation}
\label{A} A_{\bm{e}}(\omega)  = \bm{e} \cdot \int dt\, e^{-i \omega
t} \frac{d^2}{dt^2} \langle \hat{\bm{d}} \rangle (t),
\end{equation}
with $ \langle \hat{\bm{d}} \rangle (t) = \textnormal{Tr}\left[\hat{\rho}(t) \hat{\bm{d}}\right]=
\sum_{\bm{\alpha}} P_{\bm{\alpha}} \langle \Psi_{\bm{\alpha}}(t) |\hat{\bm{d}} |
 \Psi_{\bm{\alpha}}(t)
 \rangle$ the expectation value of the dipole operator $\hat{\bm d}$ of the molecule. The corresponding power density reads~\cite{MilonniD,BurnettHHG}:
\begin{equation}
\label{spectrum} S_{\bm{e}}(\omega) \propto \vert A_{\bm{e}}
(\omega) \vert^2.
\end{equation}
We note  that one can observe interferences in the
intensity $S(\omega)$ from incoherent members of the ensemble, i.e.,
members belonging to different $\bm{\alpha}$. This effect of
intensity interferences stemming from adding electric fields is
known as ``polarization beats" to distinguish it from coherent
quantum beats \cite{Faeder}.

We insert the wave function \eqref{wavefunction} into the
expression for $ \langle \hat{\bm{d}} \rangle (t)$ and obtain
\begin{equation}
\label{d} \langle \hat{\bm{d}} \rangle (t) = \int \, d \Omega \,
G(\Omega,t_p) \langle \hat{\bm{d}} \rangle^{e,\text{vib}
}(t;t_p,\Omega),
\end{equation}
with the vibronic dipole $\langle \hat{\bm{d}}
\rangle^{e,\text{vib}} (t;t_p,\Omega) = \langle \psi (t; t_p,
\Omega)\vert \hat{\bm{d}}_e \vert \psi (t; t_p, \Omega)\rangle$ and
\begin{equation}
\label{G} G(\Omega,t_p) = \sum_{JM} P_J \vert
\phi_{JM}(\Omega,t_p)\vert^2,
\end{equation}
the angular distribution at time $t_p$. The corresponding spectrum
is obtained from \eqref{A}-\eqref{spectrum}
\begin{equation}
\label{S_simple} S(\omega,t_p) \propto \left\vert \int
 d\Omega \,G(\Omega,t_p) A_{\bm{e}}^\text{$e$,vib}(\omega, \Omega, t_p) \right\vert^2,
\end{equation}
with $A_{\bm{e}}^\text{$e$,vib}(\omega,\Omega,t_p)$ the complex
amplitude for generation of high-harmonics at frequency $\omega$ and
polarization $\bm{e}$ from the electronic dipole $\langle
\hat{\bm{d}} \rangle^{e,\text{vib}} (t;t_p,\Omega) $ in a molecule
fixed at $\Omega$. If no alignment pulses are used, $G(\Omega, t_p)$
is isotropic. This follows from $G(\Omega,t_p) =
\sum_J P_J \sum_M \vert Y_{JM} (\Omega) \vert^2 = \sum_J P_J (2J+1)
/ (4 \pi)$ which is indeed independent of angles. In this case, the
spectrum arises from the coherent summation of amplitude
contributions from different orientations,
$A_{\bm{e}}^\text{$e$,vib}(\omega,\Omega,t_p)$, independent on the
temperature of the sample. In general, the signals always contain
optical coherences, except in the trivial case
$G(\Omega,t_p)=\delta(\Omega - \Omega')$. Apart from being
physically well-justified and displaying the effects of the frozen
molecular degrees of freedom, the relevance of \eqref{S_simple} is
that the numerical propagation during the ultrashort pulse is
immensely more manageable when the dimensionality is reduced. For
molecules with vibrational frequencies much smaller than the inverse
duration of the applied laser pulse, the vibrational coordinates can
also be treated as fixed, and one can separate out the vibrational
part of the vibronic wave function in \eqref{S_simple} as well.
\begin{figure}
  \begin{center}
    \includegraphics[width=\columnwidth]{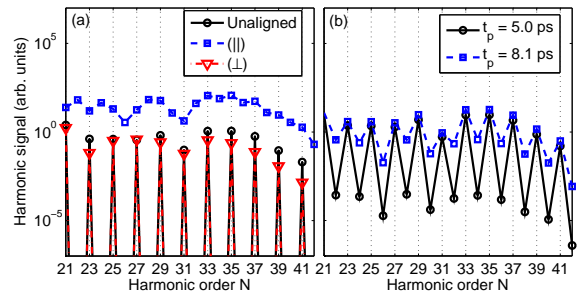}
  \end{center}
  \caption{(Color online) Harmonic generation from CO. The signal is polarized along the
linearly polarized driving laser of wave length $800$ nm and
intensity $2\times10^{14}$ W/cm$^2$. (a) HHG yield for  (i) an isotropic, unaligned ensemble, (ii) perfect
orientation along the driving laser polarization ($\parallel$) and
(iii) perfect orientation perpendicular to the driving laser
polarization ($\perp$). (b) HHG yield from CO after
 field-free orientation by a half cycle pump pulse. The
delay of the probe pulse with respect to the orienting pulse is
denoted by $t_p$. }
  \label{fig:fig1}
\end{figure}
For  CO with frozen rotations, Fig.~\ref{fig:fig1} shows the results of simulations of the
emitted high-harmonics of the same polarization as the linearly
polarized driving 800 nm probe field with peak intensity $2\times
10^{14}$ W/cm$^2$. The model used to calculate the harmonic signal
resembles the Lewenstein model. Further details are given
in~\cite{madsen:023403}.
Panel (a) shows three cases corresponding to (i) uniform
orientational distribution, (ii) perfect parallel orientation, i.e.,
the permanent molecular dipoles (each pointing from the O to
the less electronegative C) are all directed along the
polarization vector of the probe pulse and (iii) perfect
perpendicular orientation.  The panel clearly shows that the HHG signals depend critically on the direction of the orientation.
We first discuss cases (ii) and (iii) and return to the unaligned case in
the end of this section. There is no dependence of the azimuthal
angle in any of the cases, and so we only consider the polar part of
the solid angle. 
The  occurrence and absence of even harmonics in the signal 
in (ii) and (iii), respectively, may be understood 
by recalling that emission of
harmonic of order N comprises N+1 dipole transitions (N absorptions
of a laser photon, and the emission of one high-harmonic photon). In
case (ii), only the component $\hat{\parallel}$ of 
the dipole operator  parallel to the
internuclear axis is present and this component
has a $\Delta \Lambda=0$ selection rule, with $\Lambda$ the absolute value of the projection of the electronic orbital angular momentum on the internuclear axis. Initially the molecule is in its  $\Sigma$ ground state and because of $\Delta \Lambda=0$ it stays in the manifold of
$\Sigma$ states. The $\Sigma$ state, from which the final
recombination step occurs, is hence accessible by the absorption of
both an even and odd number of photons and, consequently, both even or odd
harmonics are produced. Turning to case (iii), only the
component $\hat{\perp}$ of the dipole operator perpendicular to
the internuclear axis is present and this component has a $\Delta
\Lambda=\pm 1$ selection rule. Consequently, only odd harmonics are
observed since the $\Pi$ state,  from which the
recombination occurs, can only be reached by the absorption of an odd
number of photons.  In the unaligned case (i), the situation 
is analyzed by considering the transition operator $\hat{O}_N=\Pi_{i=1}^{N+1}(\hat{\parallel}_i\cos\theta+\hat{\perp}_i\sin\theta)$
corresponding to emission of a harmonic of order N for a molecule
with the permanent dipole oriented at an angle $\theta$ with respect
to the polarization vector of the probe laser field. In the limits
of parallel ($\theta=0^\circ$) and perpendicular ($\theta=90^\circ$)
orientation we retrieve the results discussed above. 
In general the operator $\hat{O}_N$ contains even
and odd powers of cosines and sines. In the unaligned case
$G(180^\circ-\theta,t_p)=G(\theta,t_p)$, and we see from Eq.~\eqref{S_simple} that only combinations
of the cosines and sines yielding an even function on $[0, \pi]$ will survive,
i.e., the terms containing an even number of the $\hat{\parallel}_i$
operator. From the selection rules it is, however, clear that the
total number of $\Lambda$ changing transition must be an even number and
thus the total number of dipole transitions is even,  explaining why
only odd harmonics are emitted in 
the unaligned case.

In order to simulate a more
realistic orientational distribution, we present in (b) the
prediction of the harmonic signal from an ensemble of partially
oriented CO molecules. To obtain orientation, a half cycle pulse
(HCP) with amplitude $870$ kV/cm and a duration of 0.5 ps (FWHM) is
followed $4.14$ ps later by a linearly polarized laser pulse of 0.5
ps duration and a peak intensity of $5\times10^{12}$ W/cm$^2$. In order
to model the orientation we solve the time-dependent Schr\"{o}dinger
equation for the rotational degrees of freedom subject to the
rotational constant $B=57.9$ GHz, dipole moment $\mu=0.112$ D, and
polarizability volume components $\alpha_{\parallel}=1.925$
{\AA}$^3$ and $\alpha_{\perp}=1.420$ {\AA}$^3$~\cite{liao04}. The
initial rotational temperature is 5 K. The probe field which
generates high-harmonics is as in panel (a). We plot the harmonic
signal at two different delays with respect to the peak of the HCP.
The time delay $t_p=5$ ps is chosen to illustrate the case with an
almost symmetric orientational distribution:
$G(180^\circ-\theta,5~\mathrm{ps})\simeq
G(\theta,5~\mathrm{ps})$, and in this case the even harmonics are 
suppressed as expected from arguments similar to the ones used in the discussion of the unaligned case in panel (a).
At longer delays the molecules have time
to orient obtaining a maximum after $8.1$ ps with
$\langle\cos\theta\rangle=-0.11$. At this time delay the CO dipoles tend
to be pointing opposite to the polarization vector of the HCP, and this asymmetric distribution allows for even and odd harmonics of comparable strength.

We now turn to a discussion of ATI.
The fundamental quantity is the
probability $W(\bm{k})$ for measuring the momentum ${\bm{k}}$ of the
outgoing electron.  The associated measurement projection operator
 is $\hat{{\cal
P}}_{\bm k} = \vert \psi^-_{\bm{k}} \rangle \langle \psi^-_{\bm{k}}
\vert \otimes  \hat{I}_R$ which projects on an electron scattering state 
$\vert \psi^-_{\bm k} \rangle$ with asymptotic momentum ${\bm k}$  
and leaves the nuclei unaffected ($ \hat{I}_R$). Accordingly, 
$W({\bm k})= \langle \hat{{\cal P}}_{\bm k}\rangle=
\textnormal{Tr}\big[\hat{\rho}(t) \hat{{\cal P}}_{\bm k}]$ where
$\hat{\rho}(t)$ is the density matrix of the system. We evaluate the trace in the position-eigenstate
basis and obtain $W({\bm k}) = \int d {\bm r}_e \int d {\bm R}
\langle {\bm r_e}, {\bm R} \vert \hat{\rho}(t) \hat{{\cal P}}_{\bm
k} \vert {\bm r}_e, {\bm R} \rangle = \sum_{\alpha} P_\alpha
\int d {\bm R}
\vert \int d {\bm r}_e {\psi^-_{\bm k}}^\star ({\bm r}_e) \Psi_\alpha
({\bm r}_e, {\bm R}, t) \vert^2$. Introducing the wave functions
\eqref{wavefunction}, we obtain:
\begin{equation}
\label{W} W(\bm{k}, t_p) = \int d \Omega\, G(\Omega, t_p)   \int dR R^2\vert
A(\bm{k}; {\bm R}, t_p)\vert^2,
\end{equation}
with $G(\Omega, t_p)$ defined in \eqref{G} and the complex
amplitude $A(\bm{k}; {\bm R}, t_p) = \int d \bm{r}_e
{\psi^-_{\bm{k}}}^\star ( \bm{r}_e) \psi(\bm{r}_e, R, t; \Omega, t_p)$
describing the transition for frozen rotations at time $t_p$. The
time $t$ may be any time after the ionizing pulse.
In contrast to
the HHG signal \eqref{S_simple} which is optically coherent in
nuclear orientations, the signal for ATI is obtained as an
incoherent summation of contributions from different molecular
orientations. As was the case for the HHG process, the calculations
involved in the evaluation of \eqref{W} are vastly simplified if the
rotational and/or the vibrational degrees of motion are frozen
during the femtosecond pulse.
\begin{figure}
  \begin{center}
    \includegraphics[width=\columnwidth]{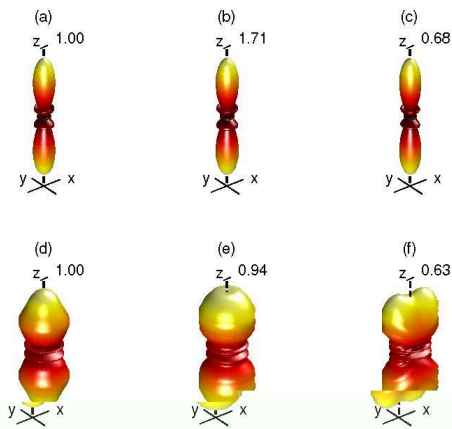}
  \end{center}
  \caption{(Color online) Angular differential ionization rates. We use an 
800 nm probe laser polarized along the $z$ axis for N$_2$ (a)-(c) at $2 \times 10^{14}$W/cm$^2$ and O$_2$ (d)-(f) at
 $1.2 \times 10^{14}$W/cm$^2$. (a) N$_2$ isotropic, unaligned ensemble.
 (b) N$_2$ maximally aligned along the $z$ axis. (c) N$_2$ maximally aligned along the $x$ axis. (d) O$_2$ isotropic, unaligned ensemble. (e) O$_2$ maximally aligned along the $z$ axis. (f) O$_2$ maximally aligned along the $x$ axis. The numbers adjacent to the $z$ axis indicate the scale with respect to the randomly aligned cases (a) and (d). The plots present an experimentally realizable demonstration of the effects of alignment on ATI.}
  \label{fig:figATI}
\end{figure}
Figure \ref{fig:figATI} shows angular distributions of the outgoing
electron for N$_2$ and O$_2$ for isotropic and field-free aligned
samples. We obtain the angular distributions by integrating
\eqref{W} over the magnitude of the momentum. In the present
calculation, we approximate the transition amplitude in
\eqref{W} by the molecular strong-field approximation
amplitude~\cite{kjeldsen:2033}. The degrees of alignment are
determined by the procedure described in \cite{ortigoso:3870}, and we use the following parallel (perpendicular)
polarizability volume of N$_2$: $2.38$ {\AA}$^3$ ($1.45$ {\AA}$^3$)
and of O$_2$: $2.3$ {\AA}$^3$ ($1.1$ {\AA}$^3$). The linearly
polarized aligning pulse has peak intensity $4 \times 10^{13}$
W/cm$^2$ and duration 59 fs (FWHM). The initial rotational
temperature is $11$ K. For the relatively low initial
rotational temperature assumed here, the maximum alignment occurs at
the quarter revival period (2.1 ps after the pump pulse for N$_2$
and 2.9 ps delay for O$_2$). The corresponding values of $\langle
\cos^2 \theta\rangle$ are 0.71 (N$_2$) and 0.74 (O$_2$). We present
results both for alignment preferentially along the polarization
axis of the probe laser and for alignment along an axis
perpendicular to the probe polarization. For N$_2$, the only effect
of making $G(\Omega,t_p)$ anisotropic is a change in the overall
scaling. This is due to the $\Sigma_g$ symmetry of the initial
orbital and is in accordance with the predictions of tunneling
theory which favors electron-ejection along the polarization
direction \cite{kjeldsen:023407}. In the case of O$_2$, we find a
more pronounced effect of the orientation. The change in angular
pattern reflects the symmetry of the initial $\Pi_g$ orbital which
has zero amplitude along and perpendicular to the molecular axis:
this nodal structure forbids the electron escaping along the
vertical polarization axis when perfectly aligned, hence the change
from (d) to (e) and (f).

In conclusion, we have developed the theory of how to deal with
frozen nuclear degrees of freedom in HHG and ATI. The frozen
coordinates affect the motion in a non-trivial way and coherence
issues depend on the degree of orientation. Physically, the phase
coherence in HHG with respect to different orientations may be
understood by the superposition principle for the electric field
generated by an ensemble of oscillators with different orientations.
The incoherence in the ATI signal follows from the fact that we in
principle, by looking at the nuclear motion after the pulse, may
infer the orientation of the molecule at the instant of ionization.
The present work form the theoretical basis for future work on
molecules interacting with strong few-cycle pulses.

This work is supported by the Danish Research Agency (Grant. No.
2117-05-0081).


\begin{thebibliography}{22}
\expandafter\ifx\csname natexlab\endcsname\relax\def\natexlab#1{#1}\fi
\expandafter\ifx\csname bibnamefont\endcsname\relax
  \def\bibnamefont#1{#1}\fi
\expandafter\ifx\csname bibfnamefont\endcsname\relax
  \def\bibfnamefont#1{#1}\fi
\expandafter\ifx\csname citenamefont\endcsname\relax
  \def\citenamefont#1{#1}\fi
\expandafter\ifx\csname url\endcsname\relax
  \def\url#1{\texttt{#1}}\fi
\expandafter\ifx\csname urlprefix\endcsname\relax\def\urlprefix{URL }\fi
\providecommand{\bibinfo}[2]{#2}
\providecommand{\eprint}[2][]{\url{#2}}

\bibitem[{\citenamefont{Itatani et~al.}(2004)\citenamefont{Itatani, Levesque,
  Zeidler, Niikura, P\'{e}pin, Kieffer, Corkum, and Villeneuve}}]{itatani04}
\bibinfo{author}{\bibfnamefont{J.}~\bibnamefont{Itatani}},
  \bibinfo{author}{\bibfnamefont{J.}~\bibnamefont{Levesque}},
  \bibinfo{author}{\bibfnamefont{D.}~\bibnamefont{Zeidler}},
  \bibinfo{author}{\bibfnamefont{H.}~\bibnamefont{Niikura}},
  \bibinfo{author}{\bibfnamefont{H.}~\bibnamefont{P\'{e}pin}},
  \bibinfo{author}{\bibfnamefont{J.~C.} \bibnamefont{Kieffer}},
  \bibinfo{author}{\bibfnamefont{P.~B.} \bibnamefont{Corkum}},
  \bibnamefont{and} \bibinfo{author}{\bibfnamefont{D.~M.}
  \bibnamefont{Villeneuve}}, \bibinfo{journal}{Nature (London)}
  \textbf{\bibinfo{volume}{432}}, \bibinfo{pages}{867} (\bibinfo{year}{2004}).

\bibitem[{\citenamefont{Niikura et~al.}(2002)\citenamefont{Niikura,
  L\'{e}gar\'{e}, Hasbani, Bandrauk, Ivanov, Villeneuve, and
  Corkum}}]{Niikura02}
\bibinfo{author}{\bibfnamefont{H.}~\bibnamefont{Niikura}},
  \bibinfo{author}{\bibfnamefont{F.}~\bibnamefont{L\'{e}gar\'{e}}},
  \bibinfo{author}{\bibfnamefont{R.}~\bibnamefont{Hasbani}},
  \bibinfo{author}{\bibfnamefont{A.~D.} \bibnamefont{Bandrauk}},
  \bibinfo{author}{\bibfnamefont{M.~Y.} \bibnamefont{Ivanov}},
  \bibinfo{author}{\bibfnamefont{D.~M.} \bibnamefont{Villeneuve}},
  \bibnamefont{and} \bibinfo{author}{\bibfnamefont{P.~B.}
  \bibnamefont{Corkum}}, \bibinfo{journal}{Nature (London)}
  \textbf{\bibinfo{volume}{417}}, \bibinfo{pages}{917} (\bibinfo{year}{2002}).

\bibitem[{\citenamefont{Niikura et~al.}(2003)\citenamefont{Niikura,
  L\'{e}gar\'{e}, Hasbani, Ivanov, Villeneuve, and Corkum}}]{Niikura03}
\bibinfo{author}{\bibfnamefont{H.}~\bibnamefont{Niikura}},
  \bibinfo{author}{\bibfnamefont{F.}~\bibnamefont{L\'{e}gar\'{e}}},
  \bibinfo{author}{\bibfnamefont{R.}~\bibnamefont{Hasbani}},
  \bibinfo{author}{\bibfnamefont{M.~Y.} \bibnamefont{Ivanov}},
  \bibinfo{author}{\bibfnamefont{D.~M.} \bibnamefont{Villeneuve}},
  \bibnamefont{and} \bibinfo{author}{\bibfnamefont{P.~B.}
  \bibnamefont{Corkum}}, \bibinfo{journal}{Nature (London)}
  \textbf{\bibinfo{volume}{421}}, \bibinfo{pages}{826} (\bibinfo{year}{2003}).

\bibitem[{\citenamefont{Alnaser et~al.}(2004)\citenamefont{Alnaser, Tong,
  Osipov, Voss, Maharjan, Ranitovic, Ulrich, Shan, Chang, Lin
  et~al.}}]{alnaser:183202}
\bibinfo{author}{\bibfnamefont{A.~S.} \bibnamefont{Alnaser}},
  \bibinfo{author}{\bibfnamefont{X.~M.} \bibnamefont{Tong}},
  \bibinfo{author}{\bibfnamefont{T.}~\bibnamefont{Osipov}},
  \bibinfo{author}{\bibfnamefont{S.}~\bibnamefont{Voss}},
  \bibinfo{author}{\bibfnamefont{C.~M.} \bibnamefont{Maharjan}},
  \bibinfo{author}{\bibfnamefont{P.}~\bibnamefont{Ranitovic}},
  \bibinfo{author}{\bibfnamefont{B.}~\bibnamefont{Ulrich}},
  \bibinfo{author}{\bibfnamefont{B.}~\bibnamefont{Shan}},
  \bibinfo{author}{\bibfnamefont{Z.}~\bibnamefont{Chang}},
  \bibinfo{author}{\bibfnamefont{C.~D.} \bibnamefont{Lin}},
  \bibnamefont{et~al.}, \bibinfo{journal}{Phys. Rev. Lett.}
  \textbf{\bibinfo{volume}{93}}, \bibinfo{eid}{183202} (\bibinfo{year}{2004}).

\bibitem[{\citenamefont{Alnaser et~al.}(2003)\citenamefont{Alnaser, Osipov,
  Benis, Wech, Shan, Cocke, Tong, and Lin}}]{alnaser:163002}
\bibinfo{author}{\bibfnamefont{A.~S.} \bibnamefont{Alnaser}},
  \bibinfo{author}{\bibfnamefont{T.}~\bibnamefont{Osipov}},
  \bibinfo{author}{\bibfnamefont{E.~P.} \bibnamefont{Benis}},
  \bibinfo{author}{\bibfnamefont{A.}~\bibnamefont{Wech}},
  \bibinfo{author}{\bibfnamefont{B.}~\bibnamefont{Shan}},
  \bibinfo{author}{\bibfnamefont{C.~L.} \bibnamefont{Cocke}},
  \bibinfo{author}{\bibfnamefont{X.~M.} \bibnamefont{Tong}}, \bibnamefont{and}
  \bibinfo{author}{\bibfnamefont{C.~D.} \bibnamefont{Lin}},
  \bibinfo{journal}{Phys. Rev. Lett.} \textbf{\bibinfo{volume}{91}},
  \bibinfo{eid}{163002} (\bibinfo{year}{2003}).

\bibitem[{\citenamefont{Baker et~al.}(2006)\citenamefont{Baker, Robinson,
  Haworth, Teng, Smith, Chirila, Lein, Tisch, and Marangos}}]{S.Baker04212006}
\bibinfo{author}{\bibfnamefont{S.}~\bibnamefont{Baker}},
  \bibinfo{author}{\bibfnamefont{J.~S.} \bibnamefont{Robinson}},
  \bibinfo{author}{\bibfnamefont{C.~A.} \bibnamefont{Haworth}},
  \bibinfo{author}{\bibfnamefont{H.}~\bibnamefont{Teng}},
  \bibinfo{author}{\bibfnamefont{R.~A.} \bibnamefont{Smith}},
  \bibinfo{author}{\bibfnamefont{C.~C.} \bibnamefont{Chirila}},
  \bibinfo{author}{\bibfnamefont{M.}~\bibnamefont{Lein}},
  \bibinfo{author}{\bibfnamefont{J.~W.~G.} \bibnamefont{Tisch}},
  \bibnamefont{and} \bibinfo{author}{\bibfnamefont{J.~P.}
  \bibnamefont{Marangos}}, \bibinfo{journal}{Science}
  \textbf{\bibinfo{volume}{312}}, \bibinfo{pages}{424} (\bibinfo{year}{2006}).

\bibitem[{\citenamefont{Paulus et~al.}(2003)\citenamefont{Paulus, Lindner,
  Walther, Baltu\v{s}ka, Goulielmakis, Lezius, and Krausz}}]{paulus:253004}
\bibinfo{author}{\bibfnamefont{G.~G.} \bibnamefont{Paulus}},
  \bibinfo{author}{\bibfnamefont{F.}~\bibnamefont{Lindner}},
  \bibinfo{author}{\bibfnamefont{H.}~\bibnamefont{Walther}},
  \bibinfo{author}{\bibfnamefont{A.}~\bibnamefont{Baltu\v{s}ka}},
  \bibinfo{author}{\bibfnamefont{E.}~\bibnamefont{Goulielmakis}},
  \bibinfo{author}{\bibfnamefont{M.}~\bibnamefont{Lezius}}, \bibnamefont{and}
  \bibinfo{author}{\bibfnamefont{F.}~\bibnamefont{Krausz}},
  \bibinfo{journal}{Phys. Rev. Lett.} \textbf{\bibinfo{volume}{91}},
  \bibinfo{eid}{253004} (\bibinfo{year}{2003}).

\bibitem[{\citenamefont{Haworth et~al.}(2007)\citenamefont{Haworth,
  Chipperfield, Robinson, Knight, Marangos, and Tisch}}]{Haworth06}
\bibinfo{author}{\bibfnamefont{C.~A.} \bibnamefont{Haworth}},
  \bibinfo{author}{\bibfnamefont{L.~E.} \bibnamefont{Chipperfield}},
  \bibinfo{author}{\bibfnamefont{J.~S.} \bibnamefont{Robinson}},
  \bibinfo{author}{\bibfnamefont{P.~L.} \bibnamefont{Knight}},
  \bibinfo{author}{\bibfnamefont{J.~P.} \bibnamefont{Marangos}},
  \bibnamefont{and} \bibinfo{author}{\bibfnamefont{J.~W.~G.}
  \bibnamefont{Tisch}}, \bibinfo{journal}{Nature Physics}
  \textbf{\bibinfo{volume}{3}}, \bibinfo{pages}{52} (\bibinfo{year}{2007}).

\bibitem[{\citenamefont{Martiny and Madsen}(2006)}]{martiny:093001}
\bibinfo{author}{\bibfnamefont{C.~P.~J.} \bibnamefont{Martiny}}
  \bibnamefont{and} \bibinfo{author}{\bibfnamefont{L.~B.}
  \bibnamefont{Madsen}}, \bibinfo{journal}{Phys. Rev. Lett.}
  \textbf{\bibinfo{volume}{97}}, \bibinfo{eid}{093001} (\bibinfo{year}{2006}).

\bibitem[{\citenamefont{Stapelfeldt and Seideman}(2003)}]{stapelfeldt03}
\bibinfo{author}{\bibfnamefont{H.}~\bibnamefont{Stapelfeldt}} \bibnamefont{and}
  \bibinfo{author}{\bibfnamefont{T.}~\bibnamefont{Seideman}},
  \bibinfo{journal}{Rev. Mod. Phys.} \textbf{\bibinfo{volume}{75}},
  \bibinfo{pages}{543} (\bibinfo{year}{2003}).

\bibitem[{\citenamefont{Hay et~al.}(2002)\citenamefont{Hay, Velotta, Lein,
  de~Nalda, Heesel, Castillejo, and Marangos}}]{PhysRevA.65.053805}
\bibinfo{author}{\bibfnamefont{N.}~\bibnamefont{Hay}},
  \bibinfo{author}{\bibfnamefont{R.}~\bibnamefont{Velotta}},
  \bibinfo{author}{\bibfnamefont{M.}~\bibnamefont{Lein}},
  \bibinfo{author}{\bibfnamefont{R.}~\bibnamefont{de~Nalda}},
  \bibinfo{author}{\bibfnamefont{E.}~\bibnamefont{Heesel}},
  \bibinfo{author}{\bibfnamefont{M.}~\bibnamefont{Castillejo}},
  \bibnamefont{and} \bibinfo{author}{\bibfnamefont{J.~P.}
  \bibnamefont{Marangos}}, \bibinfo{journal}{Phys. Rev. A}
  \textbf{\bibinfo{volume}{65}}, \bibinfo{pages}{053805}
  (\bibinfo{year}{2002}).

\bibitem[{\citenamefont{Itatani et~al.}(2005)\citenamefont{Itatani, Zeidler,
  Levesque, Spanner, Villeneuve, and Corkum}}]{itatani:123902}
\bibinfo{author}{\bibfnamefont{J.}~\bibnamefont{Itatani}},
  \bibinfo{author}{\bibfnamefont{D.}~\bibnamefont{Zeidler}},
  \bibinfo{author}{\bibfnamefont{J.}~\bibnamefont{Levesque}},
  \bibinfo{author}{\bibfnamefont{M.}~\bibnamefont{Spanner}},
  \bibinfo{author}{\bibfnamefont{D.~M.} \bibnamefont{Villeneuve}},
  \bibnamefont{and} \bibinfo{author}{\bibfnamefont{P.~B.}
  \bibnamefont{Corkum}}, \bibinfo{journal}{Phys. Rev. Lett.}
  \textbf{\bibinfo{volume}{94}}, \bibinfo{eid}{123902} (\bibinfo{year}{2005}).

\bibitem[{\citenamefont{Kanai et~al.}(2005)\citenamefont{Kanai, Minemoto, and
  Sakai}}]{Kanai05}
\bibinfo{author}{\bibfnamefont{T.}~\bibnamefont{Kanai}},
  \bibinfo{author}{\bibfnamefont{S.}~\bibnamefont{Minemoto}}, \bibnamefont{and}
  \bibinfo{author}{\bibfnamefont{H.}~\bibnamefont{Sakai}},
  \bibinfo{journal}{Nature (London)} \textbf{\bibinfo{volume}{435}},
  \bibinfo{pages}{470} (\bibinfo{year}{2005}).

\bibitem[{\citenamefont{Lein et~al.}(2005)\citenamefont{Lein, Nalda, Heesel,
  Hay, Springate, Velotta, Castillejo, Knight, and Marangos}}]{Lein05}
\bibinfo{author}{\bibfnamefont{M.}~\bibnamefont{Lein}},
  \bibinfo{author}{\bibfnamefont{R.} \bibnamefont{de~Nalda}},
  \bibinfo{author}{\bibfnamefont{E.}~\bibnamefont{Heesel}},
  \bibinfo{author}{\bibfnamefont{N.}~\bibnamefont{Hay}},
  \bibinfo{author}{\bibfnamefont{E.}~\bibnamefont{Springate}},
  \bibinfo{author}{\bibfnamefont{R.}~\bibnamefont{Velotta}},
  \bibinfo{author}{\bibfnamefont{M.}~\bibnamefont{Castillejo}},
  \bibinfo{author}{\bibfnamefont{P.~L.} \bibnamefont{Knight}},
  \bibnamefont{and} \bibinfo{author}{\bibfnamefont{J.~P.}
  \bibnamefont{Marangos}}, \bibinfo{journal}{J. Mod. Opt.}
  \textbf{\bibinfo{volume}{52}}, \bibinfo{pages}{465} (\bibinfo{year}{2005}).

\bibitem[{\citenamefont{Sundaram and Milonni}(1990)}]{MilonniD}
\bibinfo{author}{\bibfnamefont{B.}~\bibnamefont{Sundaram}} \bibnamefont{and}
  \bibinfo{author}{\bibfnamefont{P.~W.} \bibnamefont{Milonni}},
  \bibinfo{journal}{Phys. Rev. A} \textbf{\bibinfo{volume}{41}},
  \bibinfo{pages}{6571} (\bibinfo{year}{1990}).

\bibitem[{\citenamefont{Burnett et~al.}(1992)\citenamefont{Burnett, Reed,
  Cooper, and Knight}}]{BurnettHHG}
\bibinfo{author}{\bibfnamefont{K.}~\bibnamefont{Burnett}},
  \bibinfo{author}{\bibfnamefont{V.~C.} \bibnamefont{Reed}},
  \bibinfo{author}{\bibfnamefont{J.}~\bibnamefont{Cooper}}, \bibnamefont{and}
  \bibinfo{author}{\bibfnamefont{P.~L.} \bibnamefont{Knight}},
  \bibinfo{journal}{Phys. Rev. A} \textbf{\bibinfo{volume}{45}},
  \bibinfo{pages}{3347} (\bibinfo{year}{1992}).

\bibitem[{\citenamefont{Faeder et~al.}(2001)\citenamefont{Faeder, Pinkas,
  Knopp, Prior, and Tannor}}]{Faeder}
\bibinfo{author}{\bibfnamefont{J.}~\bibnamefont{Faeder}},
  \bibinfo{author}{\bibfnamefont{I.}~\bibnamefont{Pinkas}},
  \bibinfo{author}{\bibfnamefont{G.}~\bibnamefont{Knopp}},
  \bibinfo{author}{\bibfnamefont{Y.}~\bibnamefont{Prior}}, \bibnamefont{and}
  \bibinfo{author}{\bibfnamefont{D.~J.} \bibnamefont{Tannor}},
  \bibinfo{journal}{J. Chem. Phys.} \textbf{\bibinfo{volume}{115}},
  \bibinfo{pages}{8440} (\bibinfo{year}{2001}).

\bibitem[{\citenamefont{Madsen and Madsen}(2006)}]{madsen:023403}
\bibinfo{author}{\bibfnamefont{C.~B.} \bibnamefont{Madsen}} \bibnamefont{and}
  \bibinfo{author}{\bibfnamefont{L.~B.} \bibnamefont{Madsen}},
  \bibinfo{journal}{Phys. Rev. A} \textbf{\bibinfo{volume}{74}},
  \bibinfo{eid}{023403} (\bibinfo{year}{2006}).

\bibitem[{\citenamefont{Y.Liao et~al.}(2004)\citenamefont{Y.Liao, Chen, and
  Chuu}}]{liao04}
\bibinfo{author}{\bibfnamefont{Y.}~\bibnamefont{Y.Liao}},
  \bibinfo{author}{\bibfnamefont{Y.~N.} \bibnamefont{Chen}}, \bibnamefont{and}
  \bibinfo{author}{\bibfnamefont{D.~S.} \bibnamefont{Chuu}},
  \bibinfo{journal}{Phys. Rev. B} \textbf{\bibinfo{volume}{70}},
  \bibinfo{pages}{233410} (\bibinfo{year}{2004}).

\bibitem[{\citenamefont{Kjeldsen and Madsen}(2004)}]{kjeldsen:2033}
\bibinfo{author}{\bibfnamefont{T.~K.} \bibnamefont{Kjeldsen}} \bibnamefont{and}
  \bibinfo{author}{\bibfnamefont{L.~B.} \bibnamefont{Madsen}},
  \bibinfo{journal}{J. Phys. B} \textbf{\bibinfo{volume}{37}},
  \bibinfo{pages}{2033} (\bibinfo{year}{2004}).

\bibitem[{\citenamefont{Ortigoso et~al.}(1999)\citenamefont{Ortigoso,
  Rodr\'{\i}guez, Gupta, and Friedrich}}]{ortigoso:3870}
\bibinfo{author}{\bibfnamefont{J.}~\bibnamefont{Ortigoso}},
  \bibinfo{author}{\bibfnamefont{M.}~\bibnamefont{Rodr\'{\i}guez}},
  \bibinfo{author}{\bibfnamefont{M.}~\bibnamefont{Gupta}}, \bibnamefont{and}
  \bibinfo{author}{\bibfnamefont{B.}~\bibnamefont{Friedrich}},
  \bibinfo{journal}{J. Chem. Phys.} \textbf{\bibinfo{volume}{110}},
  \bibinfo{pages}{3870} (\bibinfo{year}{1999}).

\bibitem[{\citenamefont{Kjeldsen and Madsen}(2006)}]{kjeldsen:023407}
\bibinfo{author}{\bibfnamefont{T.~K.} \bibnamefont{Kjeldsen}} \bibnamefont{and}
  \bibinfo{author}{\bibfnamefont{L.~B.} \bibnamefont{Madsen}},
  \bibinfo{journal}{Phys. Rev. A} \textbf{\bibinfo{volume}{74}},
  \bibinfo{eid}{023407} (\bibinfo{year}{2006}).

\end{thebibliography}
\end{document}